\documentclass[10pt,showpacs,superscriptaddress,notitlepage,pra]{revtex4-1}

\usepackage[tbtags]{amsmath}
\usepackage{amsthm}
\usepackage{amssymb}
\usepackage{amsfonts}
\usepackage{bbm,bm,mathrsfs}
\usepackage{graphicx}
\usepackage{xcolor}
\usepackage{hyperref}
\hypersetup{colorlinks=true,breaklinks=true,pdfstartview=FitH,linkcolor=blue,citecolor=brown}

\allowdisplaybreaks

\DeclareMathOperator{\Tr}{Tr}

\theoremstyle{plain}
\newtheorem*{lemma}{Lemma}
\newtheorem{corollary}{Corollery}
\newtheorem{proposition}{Proposition}

\begin{document}

\title{On the optimal measurement for quantum discord of two-qubit states}

\author{Mingjun Shi}
\email{shmj@ustc.edu.cn}
\affiliation{Department of Modern Physics, University of Science and Technology of China, Hefei, Anhui 230026, People's Republic of
China}

\author{Chunxiao Sun}
\affiliation{Department of Modern Physics, University of Science and
Technology of China, Hefei, Anhui 230026, People's Republic of
China}

\author{Fengjian Jiang}
\affiliation{Huangshan University, Huangshan, Anhui 245021, People's Republic of China}

\author{Xinhu Yan}
\affiliation{Huangshan University, Huangshan, Anhui 245021, People's Republic of China}

\author{Jiangfeng Du}
\affiliation{Department of Modern Physics, University of Science and Technology of China, Hefei, Anhui 230026, People's Republic of China}

\begin{abstract}
We present an efficient method to solve the quantum discord of two-qubit X states exactly.
A geometric picture is used to clarify whether and when the general POVM measurement is superior to von Neumann measurement.
We show that either the von Neumann measurement or the three-element POVM measurement is optimal, and more interestingly, in the latter case the components of the postmeasurement ensemble are invariant for a class of states.
\end{abstract}


\maketitle

The notation of quantum discord was proposed in 2001 \cite{Ollivier.PhysRevLett.88.017901.2001,*Henderson.JPhysA.34.6899.2001}.
It is regarded as a measure of quantumness of correlation, even in the absence of quantum entanglement.
For ten years, many works have been devoted to the significance and application of quantum discord (see, for example,
\cite{Vedral.PhysRevLett.90.050401.2003,*Datta.PhysRevLett.100.050502.2008,
      *Shabani.PhysRevLett.102.100402.2009,*Modi.PhysRevLett.104.080501.2010,
      *Cavalcanti.PhysRevA.83.032324.2011,*Madhok.PhysRevA.83.032323.2011,
      *Streltsov.PhysRevLett.106.160401.2011,*Piani.PhysRevLett.106.220403.2011,
      *Cornelio.PhysRevLett.107.020502.2011}).
The analytical expressions for quantum discord have been obtained only in a few cases including two-qubit Bell-diagonal states \cite{Luo.PhysRevA.77.042303.2008}, rank-two states
\cite{Cen.PhysRevA.83.054101.2011,*Shi.JPA.44.415304.2011} and Gaussian state
\cite{Giorda.PhysRevLett.105.020503.2010,*Adesso.PhysRevLett.105.030501.2010}.
However there is no exact result so far for two-qubit X states (i.e., the states such that the non-zero elements of the density matrix only lie along the diagonal or skew diagonal).
In this paper we present an efficient method to solve this problem.

As we have known, the major difficulty in the calculation quantum discord is how to acquire the maximal information about one particle by measuring the other particle.
Given a bipartite state $\rho^{AB}$, perform on particle $A$ a generalized measurement, denoted by POVM $M=\{M_i\}$ with $M_i\geqslant0$ $\sum_iM_i=\mathbbm1$.
The accessible information about particle $B$ is given by
$S(\rho^B)-S(\rho^B|M)$.
Here the conditional entropy $S(\rho^B|M)=\sum_ip_iS(\rho^{B|M_i})$ is the weighted average of the states $\rho^{B|M_i}=\Tr_A[(M_i\otimes\mathbbm1)\rho^{AB}]/p_i$ that correspond to the individual outcomes with probabilities
$p_i=\Tr[(M_i\otimes\mathbbm1)\rho^{AB}]=\Tr(M_i\rho^A)$.
The maximization over all POVMs gives the classical correlation,
$\mathcal{C}=\max_{M}\big[S(\rho^B)-S(\rho^B|M)$\big].
Quantum discord $\mathcal{Q}$ is given by $\mathcal{Q}=\mathcal{I}-\mathcal{C}$, where $\mathcal{I}$ is the total correlation quantified by the mutual information, $\mathcal{I}=S(\rho^A)+S(\rho^B)-S(\rho^{AB})$.
It is a formidable task to find the optimal measurement among all $M$ to achieve the minimal value of the conditional entropy $S(\rho^B|M)$.
Much effort \cite{Ali.PhysRevA.81.042105.2010,
                       Lu.PhysRevA.83.012327.2011,
                       Girolami.PhysRevA.83.052108.2011,
                       Chen.PhysRevA.84.042313.2011},
analytical or numerical, has been made in studying the optimization for two-qubit states.
However there is no definite answer as to whether and how the quantum discord is determined by the general POVM measurements.

The measurement $M$ on $A$ induces the decomposition of $\rho^B$ into the ensemble
$\{p_i,\rho^{B|M_i}\}$.
For two-qubit states, we have known that all $\rho^{B|M_i}$ are distributed, in terms of the Bloch vectors, in an ellipsoidal region in three-dimensional real space.
This region is called quantum steering ellipsoid \cite{Verstraete.diss.2002}, which we denote by $\frak{E}$.
It has been shown that this geometric picture is very useful in the discussion of the quantum discord of two-qubit states \cite{Shi.NJP.13.073016.2011,*Shi.JPA.44.415304.2011}: We need only consider the decomposition with the form of $\rho^B=\sum_ip_i\rho^B_i$, where all $\rho^B_i$ are distributed on the surface of $\frak{E}$.
The problem is then to find the minimal value of the average entropy
$\overline{S}^B=\sum_ip_iS(\rho^B_i)$, which we denoted by
$\overline{S}^B_{\min}$.
As we show later, we benefit greatly from this geometric picture in the case of two-qubit X states: The optimal measurement on $A$, or the optimal decomposition of $\rho^B$, can be determined unambiguously, and thus the exact result of quantum discord is attained.

Note that there are infinitely many states corresponding to a given $\frak{E}$.
Denote by $[\rho^{AB}]_{\frak{E}}$ the set of all X states having the identical $\frak{E}$.
We show that all steering ellipsoids associated with X states are classified into three types. (i) $\frak{E}_{\leftrightarrow}$:
For any state in the set $[\rho^{AB}]_{\frak{E}_{\leftrightarrow}}$, the optimal measurement is such a von Neumann measurement that induces a horizontal decomposition of $\rho^B$ in the geometric picture.
(ii) $\frak{E}_{\updownarrow}$:
The optimal (von Neumann) measurement induces a vertical decomposition of $\rho^B$ for any $\rho^{AB}\in[\rho^{AB}]_{\frak{E}_{\updownarrow}}$.
(iii) $\frak{E}_{\triangle}$:
For some states in $[\rho^{AB}]_{\frak{E}_{\triangle}}$, the optimal decomposition is horizontal, while for others the three-state decomposition is optimal.
More interestingly, for the different states in the latter case, say $\rho^{AB}$ and ${\rho'}^{AB}$, the optimal decomposition is realized on the identical components, that is, the reduced states $\rho^B$ and ${\rho'}^B$ are decomposed optimally as $\sum_{i=1}^3p_i\rho_i^B$ and $\sum_{i=1}^3p'_i\rho_i^B$, respectively.

In other words, the optimal decomposition is mainly determined by the property of $\frak{E}$.
Although there is no a systematic way to describe the optimal measurement, the optimal decomposition does be described clearly in the geometric picture.

Any X state, up to local unitary operations, can be written as
\begin{equation}\label{X state}
  \rho^{AB}=
  \begin{pmatrix}
    a & 0 & 0 & u \\
    0 & b & v & 0 \\
    0 & v & c & 0 \\
    u & 0 & 0 & d
  \end{pmatrix}.
\end{equation}
where $u,v\geqslant0$ and satisfy $u^2\leqslant ad$ and $v^2\leqslant bc$.
The steering ellipsoid $\frak{E}$ is given by
$x^2/\ell_1^2+y^2/\ell_2^2+(z-z_0)^2/\ell_3^2=1$,
where $\ell_1=\frac{u+v}{\sqrt{(a+b)(c+d)}}$, $\ell_2=\frac{|u-v|}{\sqrt{(a+b)(c+d)}}$,
$\ell_3=\frac{|ad-bc|}{(a+b)(c+d)}$ and $z_0=\frac{ac-bd}{(a+b)(c+d)}$.
The reduced state $\rho^B$ corresponds to the point $B=(0,0,a-b+c-d)$, which is on the $z$ axis and in the interior of $\frak{E}$.
Note that if $u=v$ or $ad=bc$ the ellipsoid degenerates to an ellipse or a line segment.
We claim in advance that the procedure presented below can be readily applied to these degenerate cases.

\begin{figure}[thbp]
\begin{center}
  \includegraphics[width=0.9\columnwidth]{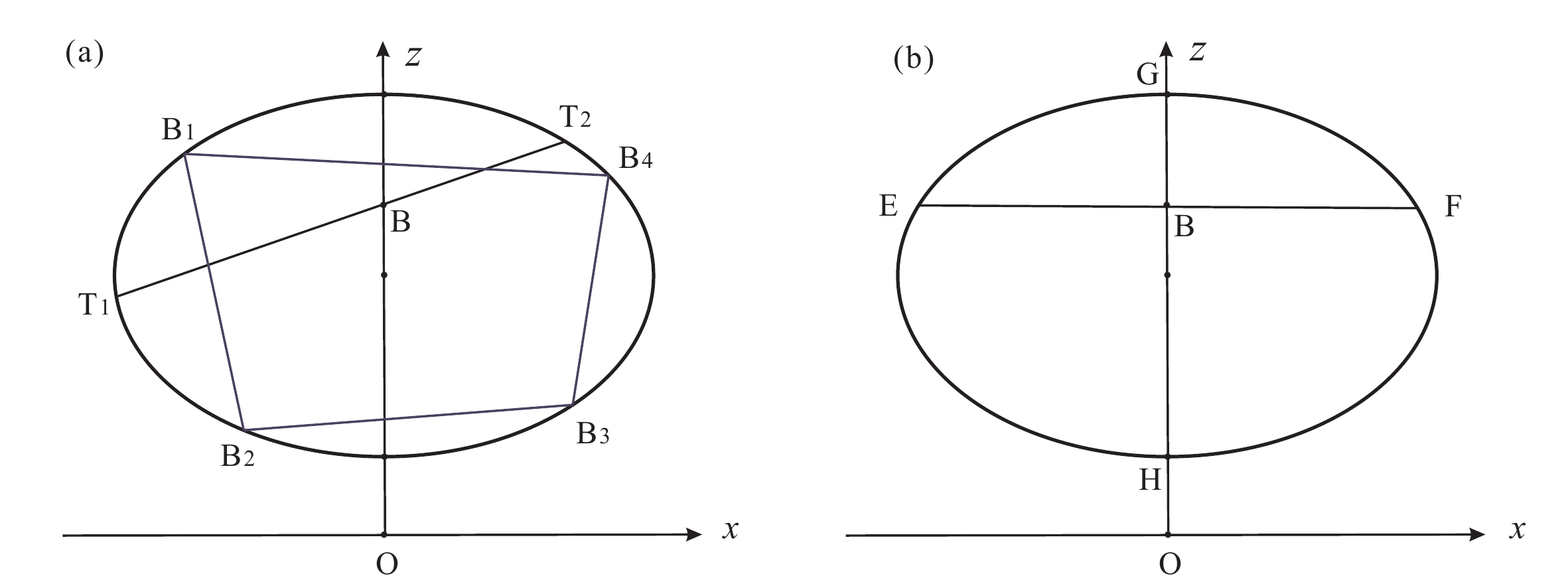}
\end{center}
\caption{Schematic plot of ellipse $\mathbf{E}$.
Point $B$ denotes the reduced state $\rho^B$.
In (a), the line segment $T_1T_2$ denotes an tilted two-state decomposition of $\rho^B$.
The inscribed quadrilateral $B_1B_2B_3B_4$ denotes a four-state decomposition of $\rho^B$.
In (b), the horizontal line segment $EF$ and the vertical $GH$ indicate the horizontal and the vertical decomposition of $\rho^B$, respectively.}\label{fig: fig1}
\end{figure}

For the state given by \eqref{X state}, the optimal von Neumann measurement must lie in the $x$-$z$ plane \cite{Chen.PhysRevA.84.042313.2011}.
It is not difficult to see that the same conclusion holds for POVM measurements:
Each operation element of the optimal POVM measurement must lie in the $x$-$z$ plane.
It means that we need only analyze the decomposition of $\rho^B$ into the convex combination of the stats on the ellipse given by
$x^2/\ell_1^2+(z-z_0)^2/\ell_3^2=1$, which we denote by $\mathbf{E}$.
All available decompositions can be described geometrically as follows.
In the ellipse $\mathbf{E}$, we plot an inscribed polygon $B_1B_2\cdots B_n$ encompassing the point $B$ (Fig. \ref{fig: fig1}(a)).
Then decomposing $\rho^B$ into $\sum_ip_i\rho_i^B$ amounts to expressing point $B$ as the convex combination of the points $B_1$, $B_2$, $\cdots$, $B_n$.
This case corresponds to the $n$-element rank-$1$ POVM measurement performed on $A$.
In fact, we need only consider the case of $n\leqslant 4$ \cite{D'Ariano.JPhysA.38.5979.2005}.
Any chord of the ellipse passing through point $B$ denotes a two-state decomposition which comes from von Neumann measurement on $A$ (see Fig. \ref{fig: fig1}(a)).

In the case that only von Neumann measurements are allowed, Ali \textit{et al} \cite{Ali.PhysRevA.81.042105.2010} claimed that for the X state \eqref{X state} the optimal observable is either $\sigma_x^A$ or $\sigma_z^A$, which give rise to the horizontal or vertical decomposition, respectively (Fig. \ref{fig: fig1}(b)).
Denote by $S_X$ the von Neumann entropy of the state corresponding to the point $X$.
The horizontal decomposition gives the average entropy
$\overline{S}^B_{\leftrightarrow}=p_ES_E+p_FS_F=S_E=S_F$, and the vertical one gives
$\overline{S}^B_{\updownarrow}=p_GS_G+p_HS_H$.
Ali's conclusion is
$\overline{S}^B_{\min}=\min\{\overline{S}^B_{\leftrightarrow},\;\overline{S}^B_{\updownarrow}\}$.
However, it is pointed out that there exist some states for which neither $\sigma_x^A$ nor $\sigma_z^A$ is optimal \cite{Lu.PhysRevA.83.012327.2011}.
A tilted decomposition, which comes from measuring the observable such as $\sigma^A_x\cos\theta+\sigma^A_z\sin\theta$, will give a smaller average entropy.
But it is not the end of the story.
Is the two-state tilted decomposition optimal? What about three- or four-state decomposition?
These problems remain open.

\begin{figure}[thbp]
\begin{center}
  \includegraphics[width=0.9\columnwidth]{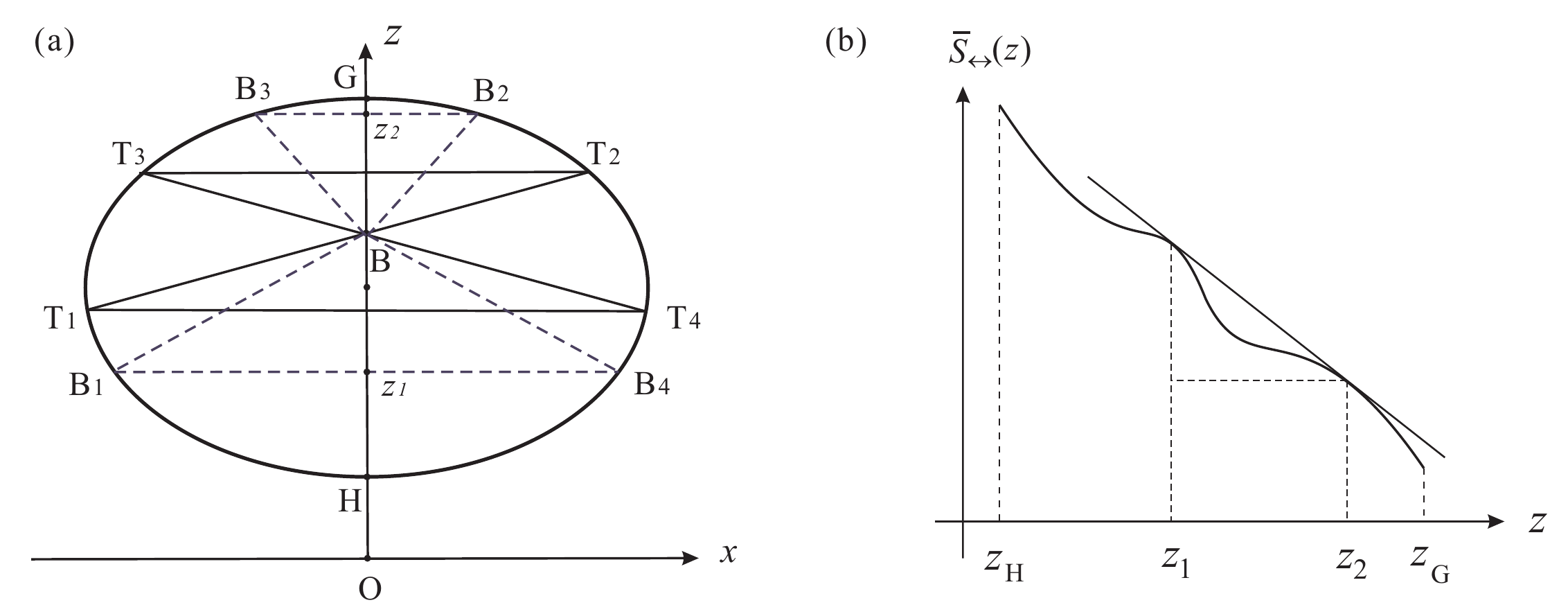}
\end{center}
\caption{In (a), the four-state decomposition $\{T_i\}_{i=1,\cdots,4}$ is equivalent to the 2-state one $\{T_1,T_2\}$, while $\{B_i\}_{i=1,\cdots,4}$ is more general.
In (b), we plot schematically the curve
$\overline{S}_{\leftrightarrow}(z)$. Two tangent points indicate the solution of \eqref{partial derivatives}.}\label{fig: fig2}
\end{figure}

To solve these problems, let us point out an evident fact.
The tilted decomposition of $\rho^B$ is equivalent to a trivial four-state decomposition.
In fact, we can plot another tilted line $T_3T_4$ (obtained by rotating $180^\circ$ around the $z$ axis) which gives rise to the same average entropy as $T_1T_2$ (see Fig. \ref{fig: fig2}(a)).
Therefore the four-state decomposition $\{T_1,T_2,T_3,T_4\}$ has the same effect as  $\{T_1,T_2\}$.
This observation leads us to the general four-state decomposition as illustrated in dashed lines in Fig. \ref{fig: fig2}(a).
Let $B_1B_3B_2B_4$ be an isosceles trapezoid inscribed in the ellipse with $B_1B_4$ and $B_2B_3$ parallel to $x$ axis.
Point $B$ is in the interior of the trapezoid.
Then the average entropy is given by $\overline{S}_4^B=\sum_{i=1}^{4}p_iS_{B_i}$.
Point $B$ may not be the intersection point of the diagonal lines of the trapezoid.

Define the function $h(x)$ as $h(x)=-\frac{1+x}{2}\log_2\frac{1+x}{2}-\frac{1-x}{2}\log_2\frac{1-x}{2}$ for $x\in[0,1]$.
We have $S_{B_i}=h(OB_i)$, where $OB_i$ is the distance from the origin point $O$ to point $B_i$.
Since $OB_1=OB_4$, $OB_2=OB_3$, $p_1=p_4$ and $p_2=p_3$, we rewrite $\overline{S}^B_4$ as $\overline{S}^{B}_4=2p_1S_{B_1}+2p_2S_{B_2}$.
Let $z_B$, $z_1$ and $z_2$ denote, respectively, the $z$-coordinate of $B$, $B_1$ and $B_2$.
Note that $S_{B_1}$ can be expressed as $\overline{S}_{\leftrightarrow}(z_1)$, which is the average entropy given by the horizontal decomposition of the state corresponding to point $(0,z_1)$ in Fig. \ref{fig: fig2}(a).
Similarly, $S_{B_2}=\overline{S}_{\leftrightarrow}(z_2)$.
With the probabilities given by
$p_1=\frac{z_2-z_B}{2(z_2-z_1)}$ and $p_2=\frac{z_B-z_1}{2(z_2-z_1)}$,
the problem is to minimize
$\overline{S}^B_{4}=2p_1\overline{S}_{\leftrightarrow}(z_1)
+2p_2\overline{S}_{\leftrightarrow}(z_2)$
over $z_1\in[z_H,z_B]$ and $z_2\in[z_B,z_G]$ with $z_H$ and $z_G$ the coordinates of the lower and the upper vertex of the ellipse $\mathbf{E}$ respectively.
Taking partial derivatives of $\overline{S}^B_4$ with respect to $z_1$ and $z_2$, we have
\begin{equation}\label{partial derivatives}
  \frac{\partial\overline{S}_{\leftrightarrow}(z_1)}{\partial z_1}=
  \frac{\partial\overline{S}_{\leftrightarrow}(z_2)}{\partial z_2}=
  \frac{\overline{S}_{\leftrightarrow}(z_1)-\overline{S}_{\leftrightarrow}(z_2)}{z_1-z_2}.
\end{equation}
It means that if there exists a four-state optimal decomposition, then the function $\overline{S}_{\leftrightarrow}(z)$ has at least two inflection points, that is, the equation
$d^2\overline{S}_{\leftrightarrow}(z)/dz^2=0$ has at least two solutions
(see Fig. \ref{fig: fig2}(b)).
This observation motivates us to investigate the properties of the horizontal average entropy $\overline{S}_{\leftrightarrow}(z)=h\big(r(z)\big)$ with
$r(z)=\big[z^2+\ell_1^2[1-(z-z_0)^2/\ell_3^2\big]^{1/2}$
for $z\in[z_H,z_G]$.
We have the following lemma.

\begin{lemma}
The horizontal average entropy $\overline{S}_{\leftrightarrow}(z)$ has at most one inflection point.
\end{lemma}

\begin{figure}[thbp]
\begin{center}
  \includegraphics[width=0.7\columnwidth]{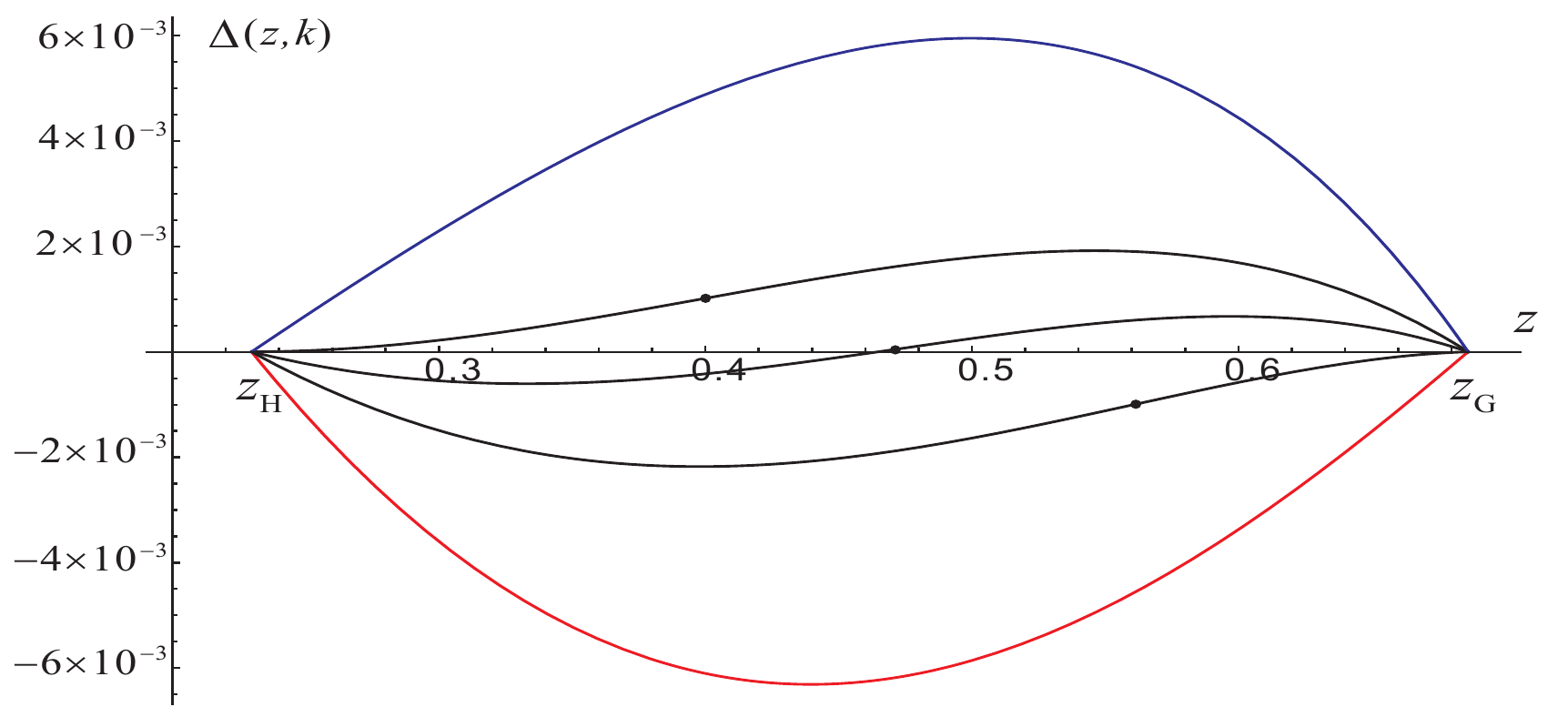}
\end{center}
\caption{(Color online) Plot of $\Delta(z,k)$ with $k$ taking the values (from bottom to top) of $0.2839$, $0.2827$, $0.2822$, $0.2817$ and $0.2805$.
The bottom curve (red) is convex, and the top (blue) is concave.
The three middle curve (black) are neither convex nor concave.
Each of them has a unique inflection point (denoted by the dot).}
\label{fig: fig3}
\end{figure}

This lemma can be proved by directly analyzing the properties of the second derivative
$d^2\,\overline{S}_{\leftrightarrow}(z)/dz^2$.
But the proof is too technical to be described here.
Instead let us consider a concrete example: a parameterized X state.
Let $\rho^{AB}(k_1,k_2)$ is such an X state that comes from \eqref{X state} with $u$ and $v$ replaced by $k_1\sqrt{ad}$ and $k_2\sqrt{ac}$ respectively.
Assume that $k_1,k_2\in[0,1]$.
When $k_1=k_2=1$, the ellipse $\mathbf{E}$ is inscribed to the unit circle.
It is the largest ellipse when $a$, $b$, $c$ and $d$ are fixed.
With $k_1$ and $k_2$ decreasing from $1$ to $0$, the ellipse shrinks to the $z$ axis while the upper and the lower vertex remain unchanged.
Note that if $k_1=k_2$, this process is similar to that of an X state undergoing local dephasing channel.
Now we let $a=0.6717$, $b=c=0.125$ and $d=0.0783$.
When $k_1=0$ and $k_2=0.8$, the state, up to local flip operations, is just the one that has been used in \cite{Lu.PhysRevA.83.012327.2011} to show that the tilted decomposition is superior to both horizontal and vertical decomposition.
In the following we let $k_1=k_2=k$ for simplicity.
For the sake of explicitness, we consider the difference $\Delta(z,k)=\overline{S}_{\leftrightarrow}(z,k)-\overline{S}_{\updownarrow}(z,k)$.
As $\overline{S}_{\updownarrow}(z,k)$ is a linear function of $z$, $\Delta(z,k)$ has the same convexity as $\overline{S}_{\leftrightarrow}(z,k)$.
In Fig. \ref{fig: fig3}, we plot $\Delta(z,k)$ for five different values of $k$.
With $k$ decreasing, we see that $\Delta(z,k)$ transforms continuously from the convex to the concave, and meanwhile there is at most one inflection point.
The same is true for $\overline{S}_{\leftrightarrow}(z)$.

Now according to Lemma, Eq. \eqref{partial derivatives} cannot be satisfied.
\begin{proposition}
  For all X states, the four-state decomposition can not be optimal.
\end{proposition}

The remaining cases are two- and three-state decomposition.
Both of them are regarded as the extremal cases of four-state decomposition.
Two-state decomposition corresponds to (i)  $z_1=z_2=z_B$ (horizontal), or (ii) $z_1=z_H$ and $z_2=z_G$ (vertical).
Three-state decomposition corresponds to (iii) $z_2=z_G$ and $z_1\in(z_H,z_B)$ (upper-triangle), or (iv) $z_1=z_H$ and $z_2\in(z_B,z_G)$ (lower-triangle).
If $\overline{S}_{\leftrightarrow}(z)$ is convex or concave, it is easy to see that case (iii) and (iv) cannot satisfy \eqref{partial derivatives}.
Then we have
\begin{proposition} \label{Prop 2}
If the horizontal average entropy $\overline{S}_{\leftrightarrow}(z)$ is a convex (concave) function for $z\in[z_H,z_G]$,
then the ellipsoid is the type of $\frak{E}_{\leftrightarrow}$ ($\frak{E}_{\updownarrow}$): for all X states associated with the ellipsoid, the optimal decomposition is the horizontal (vertical) one.
\end{proposition}

In proving the Lemma, we see that (i) if $u+v\geqslant|ad-bc|$,
$\overline{S}_{\leftrightarrow}(z)$ is a convex function, (ii) if $(u+v)^2\leqslant(a-b)(d-c)$, $\overline{S}_{\leftrightarrow}(z)$ is a concave function, and (iii) if the center of the ellipse coincides with the origin (i.e., $z_0=0$), then
$\overline{S}_{\leftrightarrow}(z)$ is either convex or concave.
Then from Proposition \ref{Prop 2} we have two corollaries.

\begin{corollary}\label{Coro 1}
  If $u+v\geqslant|ad-bc|$, the horizontal decomposition is optimal.
  If $(u+v)^2\leqslant(a-b)(d-c)$, the vertical decomposition is optimal.
\end{corollary}

\begin{corollary}\label{Coro 2}
  If the center of the ellipse coincides with the origin, then the horizontal decomposition is optimal for oblate ellipse (i.e., $\ell_1>\ell_3$), and the vertical decomposition is optimal for prolate ellipse (i.e., $\ell_1<\ell_3$).
\end{corollary}
Note that the same conclusion as Corollary \ref{Coro 1} has been obtained in \cite{Chen.PhysRevA.84.042313.2011}, and that
Corollary \ref{Coro 2} is not restricted to the states with maximally mixed marginals.

\begin{figure}[thbp]
\begin{center}
  \includegraphics[width=0.9\columnwidth]{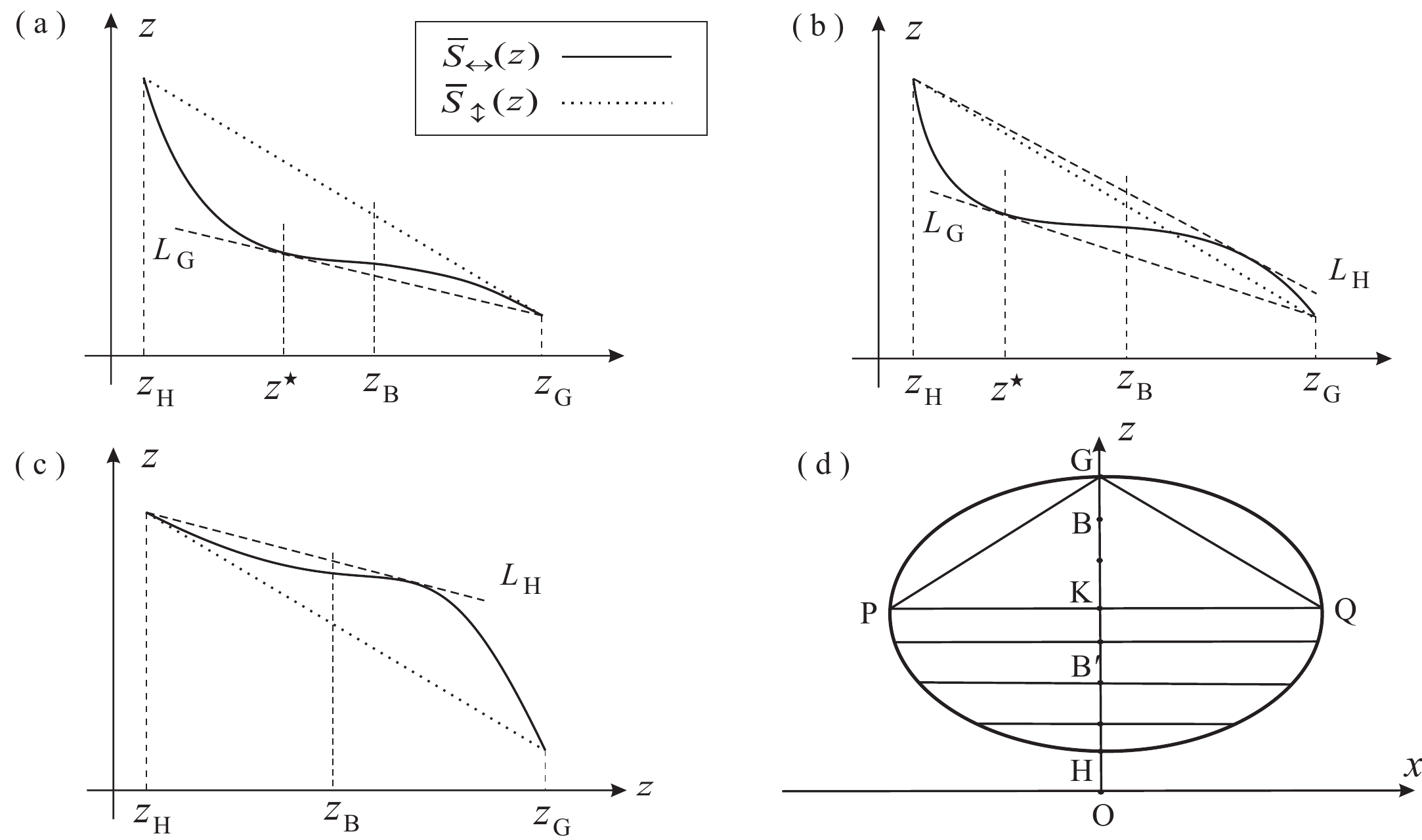}
\end{center}
\caption{Fig. (a)-(c) are the schematic plots of $\overline{S}_{\leftrightarrow}(z)$ with only one inflection point (solid curve). $\overline{S}_{\updownarrow}(z)$ is labeled by dotted line. Fig. (d) illustrates the optimal upper-triangle decomposition $\{G,P,Q\}$.}
\label{fig: fig4}
\end{figure}

Something interesting is going on here.
Let us consider the case that $\overline{S}_{\leftrightarrow}(z)$ has a unique inflection point.
The pictures of such a sort of $\overline{S}_{\leftrightarrow}(z)$ are plotted schematically in Fig. \ref{fig: fig4}(a)--(c), which in fact correspond to the three middle curves in Fig. \ref{fig: fig3}.
In the cases illustrated in Fig. \ref{fig: fig4}(a) and (b), we can always draw a line, denoted by $L_G$, tangent to the curve $\overline{S}_{\leftrightarrow}(z)$ and passing through its right end point.
Denote by $z^\star$ the $z$-coordinate of the tangent point.
We see that $z^\star$ satisfies
\begin{equation}\label{zstar}
  \frac{\partial\overline{S}_{\leftrightarrow}(z)}{\partial z}\bigg|_{z=z^\star}=
  \frac{\overline{S}_{\leftrightarrow}(z^\star)-\overline{S}_{\leftrightarrow}(z_G)}
  {z^\star-z_G}.
\end{equation}
That is, $z^\star$ is the solution of Eq. \eqref{partial derivatives} in the extremal case of $z_2=z_G$.
If $z_B\in(z^\star,z_G)$, this solution gives rise to the three-state decomposition of $\rho^B$, which leads to the minimal average entropy $\overline{S}^B_{\min}$.
Fig. \ref{fig: fig4}(d) illustrates the upper-triangle decomposition: the horizontal line $PQ$ denotes the horizontal decomposition of the point $K=(0,z^\star)$, and the optimal decomposition of $B$ is given by $\{G,P,Q\}$.
It follows that
$\overline{S}_{\min}^{B}
  =p^\star\;\overline{S}_{\leftrightarrow}(z^\star)+(1-p^\star)S_G
  =p^\star\;S_P+(1-p^\star)S_G$ with $p^\star=\frac{z_G-z_B}{z_G-z^\star}$.
We emphasize the fact that the solution $z^\star$ is independent of $z_B$.
It implies that for any point $B\in KG$, the optimal decomposition is always given by $\{G,P,Q\}$.
Of course, for $B'\in HK$, we cannot benefit from the three-state decomposition and the optimal choice for $B'$ is still the horizontal one.

In the cases illustrated in Fig. \ref{fig: fig4}(b) and (c), we can also draw a tangent line $L_H$ from the left end point.
The $z$-coordinate of the tangent point is the solution of \eqref{partial derivatives} in the extremal case of $z_1=z_H$.
This solution also provides a three-state decomposition.
But it cannot be optimal, because the average entropy $\overline{S}^B$ given by this decomposition is larger than that given by the horizontal and vertical one.

It should be mentioned that the above analysis, in particular Fig. \ref{fig: fig3} and Fig. \ref{fig: fig4}, is sensible in the case that the center of the ellipse is on the upper half of $z$ axis, namely, $z_0>0$.
If $z_0<0$, analogous analysis shows that in this case we will encounter the lower-triangle decomposition.
From these considerations, we have
\begin{proposition}\label{Prop 3}
If $\overline{S}_{\leftrightarrow}(z)$ is not larger than $\overline{S}_{\updownarrow}(z)$ for all $z\in(z_H,z_G)$, and if $\overline{S}_{\leftrightarrow}(z)$ is neither convex nor concave,
then the ellipsoid is the type of $\frak{E}_{\triangle}$:
some X states associated with the ellipsoid admit the optimal three-state decomposition.
Moreover, the components of the optimal three-state decomposition are the same for these X states.
\end{proposition}

Let us discuss some applications of our results.
Consider first the X states with the property that
$\overline{S}^B_{\leftrightarrow}=\overline{S}^B_{\updownarrow}$.
This sort of state appears on the boundary of the maximally discordant mixed states (MDMS) \cite{Galve.PhysRevA.83.012102.2011}, and also on the boundaries of the relationship between the discord and the mixedness measured by the von Neumann entropy or linear entropy \cite{AlQasimi.PhysRevA.83.032101.2011,*Girolami.JPA.44.352002.2011}.
Proposition \ref{Prop 3} shows that for these states the optimal measurement must be three-element POVM.

Another scenario in which our results are relevant is that of the sudden transition between classical and quantum decoherence presented in \cite{Maziero.PhysRevA.80.044102.2009,*Mazzola.PhysRevLett.104.200401.2010}.
It has been shown that the two-sided phase damping channel will cause the steering ellipsoid shrinking towards the $z$ axis \cite{Shi.NJP.13.073016.2011}.
The ellipse $\mathbf{E}$ transforms from an oblate ellipse (if exists initially) to a circle, then to a prolate ellipse.
Note that the states used in
\cite{Maziero.PhysRevA.80.044102.2009,*Mazzola.PhysRevLett.104.200401.2010}
to demonstrate the sudden transition are the X states with the maximally mixed marginals, which fall under the class of states with the property of $z_0=0$.
According to Corollary \ref{Coro 2}, the sudden transition follows from the sudden change of the optimal decomposition from the horizontal to the vertical.
However, for general X states with $z_0\neq0$, the optimal decomposition changes continuously, rather than suddenly, from the horizontal to the vertical.
We think that the sudden transition cannot appear in this case.

In summary, we present an efficient method to solve the quantum discord of two-qubit X states, and give a conclusive answer as to what type of and under what conditions the POVM measurement is optimal.
More interestingly, for a class of states the postmeasurement ensembles of qubit $B$ induced by the optimal three-element POVM measurement on $A$ have the identical components.
The geometric picture developed in this work is useful in discussing the boundary states such as MDMS and the dynamics of quantum discord.
The remaining question is, under what conditions is the ellipsoid the type of
$\frak{E}_{\leftrightarrow}$, $\frak{E}_{\updownarrow}$ or $\frak{E}_{\triangle}$?
Also, can this method be generalized to more general two-qubit states?
These questions will be discussed elsewhere.

This work is supported by National Nature Science Foundation of China 10875053, the CAS, and the National Fundamental Research Program 2007CB925200.

\bibliography{OnOptMeasurement}

\clearpage
\end{document}